\documentclass[doublecol]{epl2} 

\title{Localization-delocalization transition on a separatrix system of nonlinear Schr\"odinger equation with disorder}
\shorttitle{Localization-delocalization transition on a separatrix system} 

\author{A.~V.~Milovanov$^{1,}$\footnote{Also at: Department of Space Plasma Physics, Space Research Institute, Russian Academy of Sciences, Profsoyuznaya 84/32, 117997 Moscow, Russian Federation} 
and A.~Iomin$^2$}

\shortauthor{A.~V.~Milovanov and A.~Iomin}

\institute{$^1$Associazione EURATOM$-$ENEA sulla Fusione, C.~R.~Frascati, Via E. Fermi 45, C.P.-65, I-00044 Frascati, Rome, Italy, EU

$^2$Department of Physics and Solid State Institute, Technion$-$Israel Institute of Technology, Haifa, 32000, Israel}


\abstract{Localization-delocalization transition in a discrete Anderson nonlinear Schr\"odinger equation with disorder is shown to be a critical phenomenon $-$ similar to a percolation transition on a disordered lattice, with the nonlinearity parameter thought as the control parameter. In vicinity of the critical point the spreading of the wave field is subdiffusive in the limit $t\rightarrow+\infty$. The second moment grows with time as a powerlaw $\propto t^\alpha$, with $\alpha$ exactly $1/3$. This critical spreading finds its significance in some connection with the general problem of transport along separatrices of dynamical systems with many degrees of freedom and is mathematically related with a description in terms fractional derivative equations. Above the delocalization point, with the criticality effects stepping aside, we find that the transport is subdiffusive with $\alpha = 2/5$ consistently with the results from previous investigations. A threshold for unlimited spreading is calculated exactly by mapping the transport problem on a Cayley tree.}

\begin{document}

\maketitle

\section{Introduction} The problem of Anderson localization in disordered media is part of the general problem of transport of waves in disordered media. It came into focus after Anderson's suggestion that extensive disorder traps electronic wave function in the tight binding regime \cite{And}. In recent experiments with matter waves, exponential localization of a Bose-Einstein condensate released into a one-dimensional waveguide in the presence of a controlled disorder has been obtained \cite{BE}. The phenomenon is caused by destructive interference between the many scattering paths as the wave reflects off the structural inhomogeneities of the medium. An interesting new topic came with the introduction of nonlinear models \cite{Sh93} and with the realization that a weak nonlinearity of the wave process can destroy localization \cite{Sh93,PS}. Indeed it was found in direct numerical simulations of one-dimensional discrete nonlinear Schr\"odinger lattice with disorder \cite{PS,Flach} that above a certain critical strength of nonlinearity an unlimited spreading of the wave function occurs, with the second moment that grows with time as a powerlaw $\propto t^\alpha$ and the exponent $\alpha$ in the range 0.3-0.4. The exact value of $\alpha$ has remained a matter of debate. 

In this Letter we analyze delocalization processes in disordered media as a transport problem for a dynamical system with many coupled degrees of freedom. This problem is solved exactly on a Cayley tree. Our main findings are as follows. We show that the localization-delocalization transition is essentially a critical phenomenon $-$ similar to a percolation transition in random lattices. Delocalization occurs spontaneously when the strength of nonlinearity goes above a certain limit. We obtain the critical strength analytically and report it here for the first time. In vicinity of the critical point the spreading of the wave field is subdiffusive in the limit $t\rightarrow+\infty$. The second moment grows with time as a powerlaw $\propto t^\alpha$, with $\alpha$ exactly $1/3$. This critical regime is completely dominated by multi-scale long-time correlations, so that the correlation time is infinite. We describe this type of spreading in terms of a generalized diffusion equation with a fractional time the so-called Caputo fractional derivative. The critical regime finds its significance in connection with the general problem of transport along separatrices of dynamical systems with many degrees of freedom \cite{ChV}. Above criticality, nonlinear properties take a stronger role over the dynamics. With the correlations effects stepping aside we find that the transport is subdiffusive with $\alpha = 2/5$. We show that this regime corresponds to Markovian diffusion with a range-dependent diffusion coefficient consistently with the results from previous investigations \cite{Sh93,PS}. This differentiation between the two transport regimes (1/3 versus 2/5) is important at it helps to sort out some ambiguities in the reported transport exponents \cite{PS,Flach} as well as to place the various transport models on a solid mathematical background in connection with the asymptotic character of the transport.   

\section{Model} We work with a variant of discrete Anderson nonlinear Schr\"odinger equation (DANSE)
\begin{equation}
i\hbar\frac{\partial\psi_n}{\partial t} = \hat{H}_L\psi_n + \beta |\psi_n|^2 \psi_n,
\label{1} 
\end{equation}
with
\begin{equation}
\hat{H}_L\psi_n = \varepsilon_n\psi_n + V (\psi_{n+1} + \psi_{n-1}).
\label{2} 
\end{equation}
Here, $\hat H_L$ is the Hamiltonian of a linear problem in the tight binding approximation; $\beta$ characterizes the strength of nonlinearity; on-site energies $\varepsilon_n$ are randomly distributed with zero mean across a finite energy range; $V$ is hopping matrix element; and the total probability is normalized to $\sum_n |\psi_n|^2 = 1$. In what follows, $\hbar = 1$ for simplicity. For $\beta\rightarrow 0$, DANSE~(\ref{1}) reduces to the original Anderson model \cite{And}. All eigenstates are exponentially localized in that limit. We aim to understand the asymptotic ($t\rightarrow+\infty$) spreading of initially localized wave packet under the action of nonlinear term.    

\section{Analysis} For each node $n$, we define a basis of linearly localized modes, the eigenfunctions of the linear Hamiltonian $\hat H_L$. By doing so we generate a mapping of the wave number space into a functional space. For strong disorder, dimensionality of this space is infinite (countable). We consider this space as embedding space for the dynamics.\footnote{To visualize, think of each node as comprising a countable number of compactified dimensions that are expanded to form a functional space.} If we introduce an inner product: $\psi^{\prime} _{n} \circ \psi^{\prime\prime} _{n} = \sum_n (\psi_n^\prime)^* \psi^{\prime\prime} _{n}$, we obtain a Hilbert space, in which an orthogonal basis can be defined. Here, the star denotes complex conjugate. Denoting the basis modes as $\phi_{n,m}$, we have, with $\omega_m$ the eigenvalues of the linear problem, $\hat H_L \phi_{n,m} = \omega_m \phi_{n,m}$. Orthogonality implies that $\sum _n \phi^*_{n,m}\phi_{n,k} = \delta_{m,k}$, where $\delta_{m,k}$ is Kronecker's delta. In the basis of linearly localized modes the wave function $\psi_n$ can be represented as 
\begin{equation}
\psi_n = \sum_m \sigma_m (t) \phi_{n,m},
\label{3} 
\end{equation}
where we have introduced $\sigma_m (t)$, a set of complex amplitudes which describe the evolution of the wave field in time. The total probability being equal to 1 implies that $\sum_m \sigma_m^* (t)\sigma_m (t) = 1$. We now obtain a set of dynamical equations for $\sigma_m (t)$. For this, substitute~(\ref{3}) into DANSE~(\ref{1}), then multiply the both sides by $\phi^*_{n,m}$, and sum over $n$, remembering that the modes are orthogonal. The result reads  
\begin{equation}
i\dot{\sigma}_k - \omega_k \sigma_k = \beta \sum_{m_1, m_2, m_3} V_{k, m_1, m_2, m_3} \sigma_{m_1} \sigma^*_{m_2} \sigma_{m_3},
\label{4} 
\end{equation}
where the coefficients $V_{k, m_1, m_2, m_3}$ are given by
\begin{equation}
V_{k, m_1, m_2, m_3} = \sum_{n} \phi^*_{n,k}\phi_{n,m_1}\phi^*_{n,m_2}\phi_{n,m_3},
\label{5} 
\end{equation}
and we have used dot to denote time derivative. Equations~(\ref{4}) correspond to a system of coupled nonlinear oscillators with the Hamiltonian 
\begin{equation}
\hat H = \hat H_{0} + \hat H_{\rm int}, \ \ \ \hat H_0 = \sum_k \omega_k \sigma^*_k \sigma_k,
\label{6} 
\end{equation}
\begin{equation}
\hat H_{\rm int} = \frac{\beta}{2} \sum_{k, m_1, m_2, m_3} V_{k, m_1, m_2, m_3} \sigma^*_k \sigma_{m_1} \sigma^*_{m_2} \sigma_{m_3}.
\label{6+} 
\end{equation}
Here, $\hat H_{0}$ is the Hamiltonian of non-interacting harmonic oscillators and $\hat H_{\rm int}$ is the interaction Hamiltonian.\footnote{We include self-interactions into $\hat H_{\rm int}$.} Each nonlinear oscillator with the Hamiltonian   
\begin{equation}
\hat h_{k} = \omega_k \sigma^*_k \sigma_k + \frac{\beta}{2} V_{k, k, k, k} \sigma^*_k \sigma_{k} \sigma^*_{k} \sigma_{k}
\label{6+h} 
\end{equation}
and the equation of motion 
\begin{equation}
i\dot{\sigma}_k - \omega_k \sigma_k - \beta V_{k, k, k, k} \sigma_{k} \sigma^*_{k} \sigma_{k} = 0
\label{eq} 
\end{equation}
represents one nonlinear eigenstate in the system $-$ identified by its wave number $k$, unperturbed frequency $\omega_k$, and nonlinear frequency shift $\Delta \omega_{k} = \beta V_{k, k, k, k} \sigma_{k} \sigma^*_{k}$. Non-diagonal elements $V_{k, m_1, m_2, m_3}$ characterize couplings between each four eigenstates with wave numbers $k$, $m_1$, $m_2$, and $m_3$. It is understood that the excitation of each eigenstate is not other than the spreading of the wave field in wave number space. Resonances occur between the eigenfrequencies $\omega_k$ and the frequencies posed by the nonlinear interaction terms. We have,\footnote{Conditions for nonlinear resonance are obtained by accounting for the nonlinear frequency shift.} 
\begin{equation}
\omega_k = \omega_{m_1} - \omega_{m_2} + \omega_{m_3}.
\label{Res} 
\end{equation}
When the resonances happen to overlap, a phase trajectory may occasionally switch from one resonance to another. As Chirikov realized \cite{Chirikov}, any overlap of resonances will introduce a random element to the dynamics along with some transport in phase space. Applying this argument to DANSE~(\ref{1}), one sees that destruction of Anderson localization is limited to a set of resonances in a Hamiltonian system of coupled nonlinear oscillators, eqs.~(\ref{6}) and~(\ref{6+}), permitting a connected escape path to infinity. 

At this point, the focus is on topology of the random motions in phase space. We address an idealized situation first where the overlapping resonances densely fill the phase space. This is fully developed chaos, a regime that has been widely studied and discussed in the literature (e.g., Refs. \cite{ZaslavskyUFN,Sagdeev}). Concerns raised over this regime when applied to eqs.~(\ref{6}) and~(\ref{6+}) come from the fact that it requires a diverging free energy reservoir in systems with a large number of interacting degrees of freedom. Yet, developed chaos offers a simple toy-model for the transport as it corresponds with a well-understood, diffusive behavior. 

A more general, as well as more intricate, situation occurs when the random motions coexist along with regular (KAM regime) dynamics. If one takes this idea to its extreme, one ends up with the general problem of transport along separatrices of dynamical systems. This problem constitutes a fascinating nonlinear problem that has as much appeal to the mathematician as to the physicist. An original important promotion of this problem to large systems is due to Chirikov and Vecheslavov \cite{ChV}.

This type of problem occurs for slow frequencies. One finds \cite{PRE01,PRE09} that resonance-overlap conditions are satisfied along the ``percolating" orbits or separatrices of the random potential where the orbital periods diverge. The available phase space for the random dynamics can be very ``narrow" in that case. In large systems, the set of separatrices can moreover be geometrically very complex and strongly shaped. Often it can be envisaged as a fractal network at percolation as for instance in random fields with sign-symmetry \cite{PRE00}. 

There is a fundamental difference between the above two transport regimes (chaotic versus near-separatrix). The former regime is associated with an exponential loss of correlation permitting a Fokker-Planck description in the limit $t\rightarrow+\infty$. The latter regime when considered for large systems is associated with an algebraic loss of correlation, implying that the correlation time is infinite. There is no a conventional Fokker-Planck equation here, unless generalized to fractional derivatives \cite{Report,Klafter}, nor the familiar Markovian property (i.e., that the dynamics are memoryless). On the contrary, there is an interesting interplay \cite{PRE09} between randomness, fractality, and correlation, which is manifest in the fact that all Lyapunov exponents vanish in the thermodynamic limit, despite that the dynamics are intrinsically random. 

This situation of random non-chaotic dynamics with zero Lyapunov exponents, being in fact very general \cite{JMPB,PD2004}, has come to be known as ``pseudochaos." One might think of pseudochaos as occurring ``at the edge" of stochasticity and chaos, thus separating fully developed chaos from domains with regular motions. In what follows, we discuss the implications of chaotic and pseudochaotic transport for the spreading of wave field in eqs.~(4). 
  
{\it Chaotic case.} $-$ As the time correlations vanish exponentially fast, equation~(4) can conveniently be considered as a Langevin equation with the nonlinear interaction term thought as a white noise term in the limit $t\rightarrow+\infty$. There is a well-defined diffusion coefficient here, which behaves as modulus square of the complex interaction amplitude. The cubic interaction in eq.~(\ref{4}) implies that $D\propto |\sigma_n| ^6$. If the field is spread over $\Delta n$ sites, then the conservation of the probability dictates $|\sigma_n| ^2 \sim 1 / \Delta n$, leading to $D\propto 1 / (\Delta n) ^3$. One sees that the transport problem in the chaotic case is essentially a diffusion problem with a range-dependent diffusion coefficient. This range-dependence is an inverse powerlaw, as one would indeed expect for a homogeneous random system. 

Let $f = f (t, \Delta n)$ be the probability density to find an initially localized wave field at time $t$ at distance $\Delta n$ from the origin. The diffusive character of the spreading justifies 
\begin{equation}
\frac{\partial f (t, \Delta n)}{\partial t} = \frac{\partial}{\partial \Delta n}\left[W \frac{1}{(\Delta n)^{3}}\frac{\partial f (t, \Delta n)}{\partial \Delta n} \right],
\label{7} 
\end{equation}
where $W$ is a constant coefficient which collects all relevant parameters of the diffusion process. The fundamental solution or Green's function of eq.~(\ref{7}) reads
\begin{equation}
f (t, \Delta n) = \frac{5}{\Gamma (1/5)} \frac{1}{(25Wt)^{1/5}} \exp\left[-\frac{(\Delta n)^5}{25Wt}\right],
\label{8} 
\end{equation}
where the normalization $\int_0^\infty f (t, \Delta n) d\Delta n = 1$ has been applied. It is noticed that the distribution in eq.~(\ref{8}) is essentially non-Gaussian. From eq.~(\ref{8}) one immediately obtains that 
\begin{equation}
\langle (\Delta n) ^2 (t) \rangle = (25Wt)^{2/5} \Gamma (3/5) / \Gamma (1/5),
\label{8+} 
\end{equation}
where the angle brackets denote ensemble average, and $\Gamma$ is the Euler gamma function. The net result is $\langle (\Delta n) ^2 (t) \rangle \propto t^{2/5}$, consistently with the finding of Refs. \cite{Sh93,PS}.

{\it Pseudochaotic case.} $-$ This regimes takes eqs.~(\ref{4}) to an opposite extreme limit where each oscillator can only communicate with the rest of the wave field via a nearest-neighbor rule. This is a marginal regime yet permitting an escape path to infinity. Clearly, the number of coupling links is minimized in that case. When summing on the right-hand-side, the only combinations to be kept are, for the reasons of symmetry, $\sigma_{k} \sigma^*_{k} \sigma_{k}$ and $\sigma_{k-1} \sigma^*_{k} \sigma_{k+1}$. We have
\begin{equation}
i\dot{\sigma}_k - \omega_k \sigma_k = \beta V_{k} \sigma_{k} \sigma^*_{k} \sigma_{k} + 2\beta V_k^\pm \sigma_{k-1} \sigma^*_{k} \sigma_{k+1},
\label{9} 
\end{equation} 
where we have also denoted for simplicity $V_k = V_{k, k, k, k}$ and $V_k^\pm = V_{k, k-1, k, k+1}$. Equations~(\ref{9}) define an infinite ($k= 1,2,\dots$) chain of coupled nonlinear oscillators where all couplings are local (nearest-neighbor-like). The interaction Hamiltonian in eq.~(\ref{6+}) is simplified to   
\begin{equation}
\hat H_{\rm int} = \frac{\beta}{2} \sum_{k} V_{k} \sigma^*_k \sigma_{k} \sigma^*_{k} \sigma_{k} + {\beta}\sum_{k} V_k^\pm \sigma^*_k \sigma_{k-1} \sigma^*_{k} \sigma_{k+1}.
\label{6++} 
\end{equation}
We are now in position to introduce a simple lattice model for the transport. The key step is to observe that eqs.~(\ref{9}) can be mapped on a Cayley tree where each node is connected to $z=3$ neighbors (here, $z$ is the coordination number). The mapping is defined as follows. A node with coordinate $k$ represents a nonlinear eigenstate, or nonlinear oscillator with the equation of motion~(\ref{eq}). There are $z=3$ branches at each node: one that we consider ingoing represents the complex amplitude $\sigma^*_{k}$, and the other two, the outgoing branches, represent the complex amplitudes $\sigma_{k-1}$ and $\sigma_{k+1}$ respectively. These settings are schematically illustrated in Fig.~1.

\begin{figure}
\includegraphics[width=0.49\textwidth]{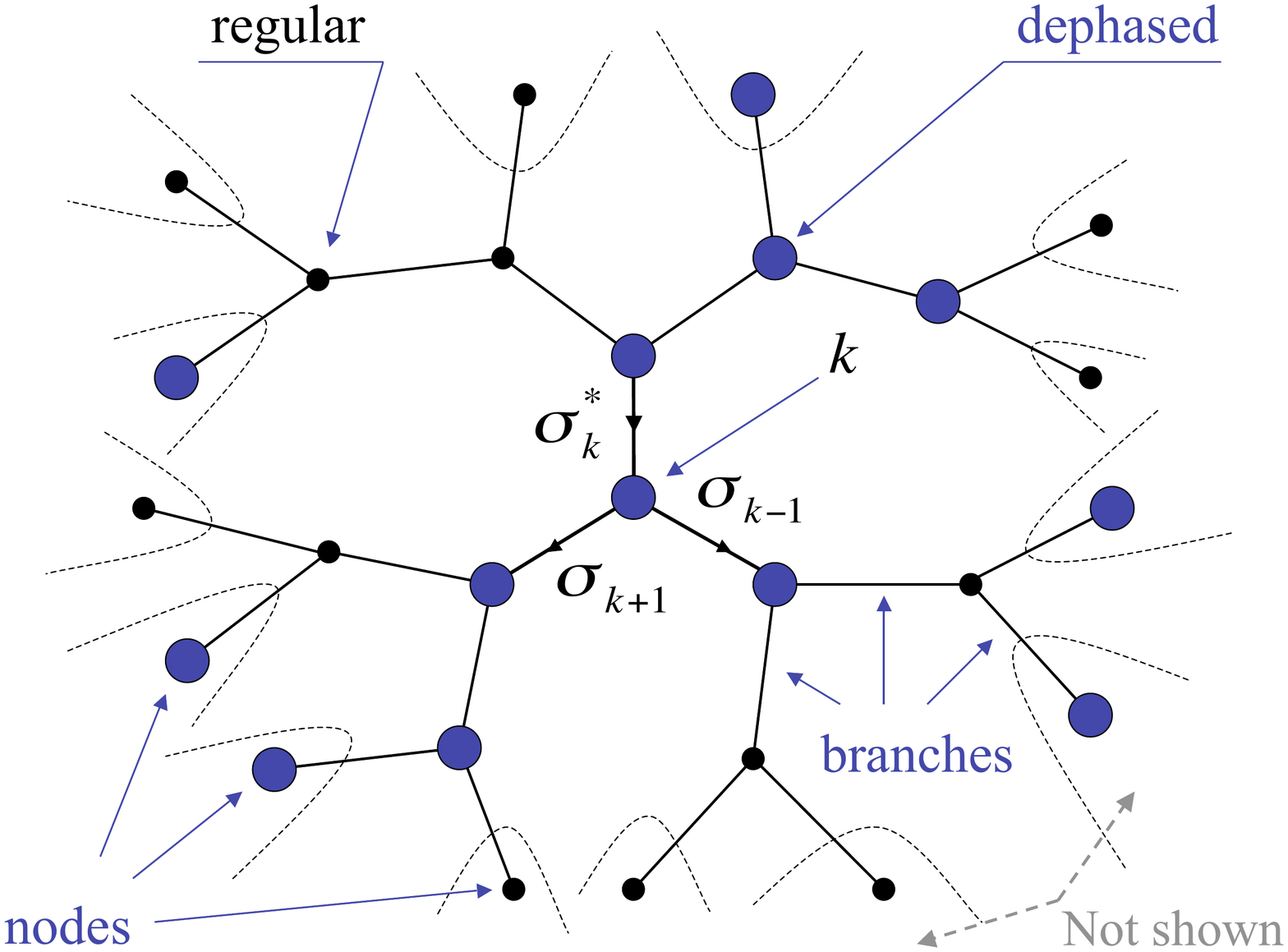}
\caption{\label{} Mapping eqs.~(\ref{9}) on a Cayley tree. Each node represents a nonlinear eigenstate, or nonlinear oscillator with the equation of motion $i\dot{\sigma}_k - \omega_k \sigma_k - \beta V_{k, k, k, k} \sigma_{k} \sigma^*_{k} \sigma_{k} = 0$. Blue nodes represent oscillators in a chaotic (``dephased") state. Black nodes represent oscillators in regular state. One ingoing and two outgoing branches on node $k$ ($k= 1,2,\dots$) represent respectively the complex amplitudes $\sigma^*_{k}$, $\sigma_{k-1}$, and $\sigma_{k+1}$. Structures that are not explicitly shown are beyond the dashed lines.}
\end{figure}

A Cayley tree being by its definition \cite{Schroeder} a hierarchical graph offers a suitable geometric model for infinite-dimensional spaces. We think of this graph as embedded into phase space of the Hamiltonian system of coupled nonlinear oscillators, eqs.~(\ref{6}) and~(\ref{6+}). In the thermodynamic limit $k_{\max}\rightarrow\infty$, in place of a Cayley tree, one uses the notion of a Bethe lattice.\footnote{A Bethe lattice is an infinite version of the Cayley tree. To this end, a purist might prefer to say ``bond" in place of ``branch," but that's all about the terminology.} Setting $k_{\max}\rightarrow\infty$, we suppose that each node of the Bethe lattice hosts a nonlinear oscillator, eq.~(\ref{eq}). The bonds of the lattice, in their turn, can conduct oscillatory processes to their neighbors as a result of the interactions present. 

Next, we assume that each oscillator can be in a chaotic (``dephased") state with the probability $p$ (and hence, in a regular state with the probability $1-p$). The $p$ value being smaller than 1 implies that the domains of random motions occupy only a fraction of the lattice nodes. Whether an oscillator is dephased is decided by Chirikov's resonance-overlap condition $-$ which may or may not be matched on node $k$. We believe that in systems with many coupled degrees of freedom each such ``decision" is essentially a matter of the probability. The choice is random. Focusing on the $p$ value, we consider system-average nonlinear frequency shift 
\begin{equation}
\Delta\omega_{\rm NL} = \beta\langle|\psi_n|^2\rangle_{\Delta n}
\label{Shift} 
\end{equation}
as an effective ``temperature" of nonlinear interaction. It is this ``temperature" that rules over the excitation of the various resonant ``levels" in the system. With this interpretation in mind, we write $p$ as the Boltzmann factor 
\begin{equation}
p = \exp (-\delta\omega / \Delta\omega_{\rm NL}),
\label{B} 
\end{equation}
where $\delta\omega$ is the characteristic energy gap between the resonances. Expanding $\psi_n$ over the basis of linearly localized modes, we have 
\begin{equation}
\langle|\psi_n|^2\rangle_{\Delta n} = \frac{1}{\Delta n} \sum_n \sum_{m_1,m_2} \phi^*_{n,m_1}\phi_{n,m_2} \sigma^*_{m_1} \sigma_{m_2}.
\label{Mean} 
\end{equation}  
The summation here is performed with the use of orthogonality of the basis modes. Combining with eq.~(\ref{Shift}),
\begin{equation}
\Delta\omega_{\rm NL} =  \frac{\beta}{\Delta n}\sum_m \sigma^*_{m} \sigma_{m}.
\label{Upon} 
\end{equation}
The sum over $m$ is easily seen to be equal to 1 due to the conservation of the probability. Thus, $\Delta\omega_{\rm NL} = \beta / \Delta n$. If the field is spread over $\Delta n$ states, then the distance between the resonant frequencies behaves as $\delta \omega \sim 1/\Delta n$. We normalize units in eq.~(\ref{1}) to have $\delta \omega = 1/\Delta n$ exactly. One sees that $p = \exp (-1/\beta)$. This result suggests that behavior be non-perturbative in the pseudochaotic regime. For the vanishing $\beta\rightarrow 0$, the Boltzmann factor $p\rightarrow 0$, implying that all oscillators are in regular state. In the opposite regime of $\beta\rightarrow\infty$, $p\rightarrow 1$. That means that all oscillators are dephased and that the random motions span the entire lattice. 

There is a critical concentration, $p_c$, of dephased oscillators permitting an escape path to infinity for the first time. This critical concentration is not other than the percolation threshold on a Cayley tree. In the basic theory of percolation it is found that $p_c = 1/(z-1)$ (e.g., Ref.~\cite{Schroeder}). This is an exact result. For $z=3$, $p_c = 1/2$. We associate the critical value $p_c = 1/2$ with the onset of transport in the DANSE model, eq.~(\ref{1}). When translated into the $\beta$ values the threshold condition reads 
\begin{equation}
\beta_c = 1/\ln (z-1).
\label{Betac} 
\end{equation}
Setting $z=3$, we have $\beta_c = 1/\ln 2 \approx 1.4427$. This value defines the critical strength of nonlinearity that destroys the Anderson localization. For the $\beta$ values smaller than this, the localization persists, despite that the problem is nonlinear. When $\beta \geq 1/\ln 2$, the localization is lost, and the wave field spreads to infinity. 

Our conclusion so far is that the loss of localization is a threshold phenomenon, which requires the strength of nonlinearity be above a certain level. In this respect, the nonlinearity parameter $\beta$ acquires the role of the control parameter. The onset of unlimited spreading is at $\beta_c = 1/\ln 2$, this being an exact result of the model. We now turn to predict second moments for the onset spreading.

This task is essentially simplified if one visualizes the transport as a random walk over a system of dephased oscillators. For $p\rightarrow p_c$, this system is self-similar, i.e., fractal. That means that dephased oscillators form arbitrarily large clusters, each presenting the same fractal geometry of the infinite percolation cluster \cite{Havlin}. It is the infinite cluster that conducts unlimited spreading of the wave function on a Bethe lattice. We should stress that the fractal geometry of the clusters is a consequence of the probabilistic character of dephasing.

In random walks on percolation systems one writes the mean-square displacement from the origin as \cite{Gefen}
\begin{equation}
\langle (\Delta n) ^2 (t) \rangle = A^2 t^{2 / (2+\theta)}, \ \ \ t\rightarrow+\infty.
\label{MS} 
\end{equation}
Here, $A$ is a dimensional constant parameter and $\theta$ is the exponent of anomalous diffusion, or the connectivity exponent. This exponent accounts for the deviation from the usual Fickian diffusion in fractal geometry. As a rule, $\theta \geq 0$, implying that the diffusion is slowed down in complex ``labyrinths" of the fractal. Another common way of writing eq.~(\ref{MS}) is given by \cite{Havlin} 
\begin{equation}
\langle (\Delta n) ^2 (t) \rangle = A^2 t^{d_s / d_f}, \ \ \ t\rightarrow+\infty,
\label{MS+} 
\end{equation}
where $d_f$ is the Hausdorff dimension which measures the number of nodes that belong to a given cluster, and $d_s = 2d_f / (2+\theta)$ is the fracton, or spectral, dimension which describes the density of states in fractal geometry. The key difference between the two dimensions is that $d_f$ is a purely structural characteristic of the fractal, whereas $d_s$ reflects the dynamical properties such as wave excitation, diffusion, etc. Note, also, that the spectral dimension is not larger than $d_f$. 

The two scaling laws above apply to any percolation system. For percolation on a Cayley tree, the following exact results hold \cite{Havlin,Naka}: $\theta =4$, $d_f = 4$, and $d_s = 4/3$. One sees that 
\begin{equation}
\langle (\Delta n) ^2 (t) \rangle \propto t^{1/3}, \ \ \ t\rightarrow+\infty.
\label{MS++} 
\end{equation}
This is the desired scaling. By its derivation, subdiffusion in eq.~(\ref{MS++}) is asymptotic ($t\rightarrow+\infty$) in the thermodynamic limit. 

We proceed with a remark that the Hausdorff dimension being equal to 4 matches with the implication of eqs.~(\ref{4}) and~(\ref{5}) where the coefficients $V_{k, m_1, m_2, m_3}$ are supposed to run over 4-dimensional subsets of the ambient Hilbert space. Indeed it is the overlap integral of four Anderson eigenmodes, eq.~(\ref{5}), that decides on dimensionality of subsets of phase space where the transport processes concentrate. When the nearest-neighbor rule is applied, this overlap structure is singled out for the dynamics. Under the condition that the structure is critical, i.e., ``at the edge" of permitting a path to infinity, the support for the transport is reduced to a percolation cluster on a Bethe lattice $-$ characterized, along with the above value of the Hausdorff dimension, by the very specific connectivity exponent, $\theta = 4$. The end result is $\alpha = 2 / (2+\theta)  = 1/3$. 

At contrast with eq.~(\ref{7}), random walks on percolation systems are described by a non-Markovian diffusion equation \cite{PRE09,NJP}: 
\begin{equation}
\frac{\partial f (t, \Delta n)}{\partial t} = \frac{1}{\Gamma_\theta}\int _{0}^{t} \frac{dt^{\prime}}{(t - t^{\prime})^{\theta / (2+\theta)}} \frac{\partial}{\partial t^{\prime}} \frac{\partial^2 f (t^\prime, \Delta n)}{\partial (\Delta n)^2},\label{23} 
\end{equation}
where $\Gamma_\theta = \Gamma (2/(2+\theta))$.
In writing eq.~(\ref{23}) we have adopted results of Ref.~\cite{PRE09} to random walks on a single cluster. The convolution on the right hand side is expressible in the compact form of a fractional time the so-called Caputo fractional derivative~\cite{Podlubny} of order $0 \leq \theta / (2+\theta) < 1$. We have, in the commonly used notations,
\begin{equation}
\frac{\partial f (t, \Delta n)}{\partial t} = {_0^c}D_t^{\theta / (2+\theta)} \frac{\partial^2 f (t, \Delta n)}{\partial (\Delta n)^2}.\label{24} 
\end{equation}
Allowing for $\theta = 4$, one sees that the critical spreading corresponds to a fractional operator ${_0^c}D_t^{2/3} = {_0^c}D_t^{1-\alpha}$ for $\alpha = 1/3$. 

The latter result should be addressed. Indeed eq.~(\ref{24}) shows that the critical spreading is a matter of {fractional}, or ``strange," kinetics \cite{Nature}, consistently with the implication of pseudochaotic behavior \cite{Report,JMPB,PD2004}. In many ways equations built on fractional derivatives offer an elegant and powerful tool to describe anomalous transport in complex systems. There is an insightful connection with a generalized master equation formalism along with a mathematically convenient way for calculating transport moments as well as solving initial and boundary value problems \cite{Klafter,Rest}. The fundamental solution of the fractional eq.~(\ref{24}) is evidenced in Table~1 of Ref.~\cite{Rest}. It shares non-Gaussianity with Green's function in eq.~(\ref{8}), being in the rest analytically very different.  

Recently, an application of the fractional diffusion equation to subdiffusion in the nonlinear Schr\"odinger equation with disorder has been proposed within a continuous time random walk (CTRW) formalism \cite{Iomin}. It was shown that there is a heavy-tailed distribution of waiting times between consecutive steps of the random motion, thus giving rise to a slower-then-linear growth of second moments. In the above eqs.~(\ref{MS}) and (\ref{MS++}) we have not as a matter of fact assumed any heavy-tailed distribution of this sort. At contrast, in the present model, the random walker is supposed to take one unit step along the cluster as soon as one unit time is elapsed. Despite this difference, the two approaches are essentially equivalent and do lead to the same type of fractional diffusion equation, eq.~(\ref{24}). This is because fractal labyrinths of the subset that hosts the random dynamics act as to introduce multi-scale trappings to the motion characterized by an inverse powerlaw-like distribution of waiting times. The latter is found to be, in properly normalized units, 
\begin{equation}
\phi (\tau) \propto 1/(1+\tau)^{(4+\theta)/(2+\theta)}.\label{CTRW} 
\end{equation}
As mean effective waiting time diverges, the basic assumptions of CTRW's are readily installed.  
        
\section{Summary} We have shown that the Anderson localization in disordered media can be lost in the presence of a weak nonlinearity and that the phenomenon is critical (thresholded). That means that there is a critical strength of nonlinearity above which the wave field turns to an unlimited spreading. Below that limit, the field is localized similarly to the linear case. We have discussed the problem as a percolation problem on a separatrix system of discrete nonlinear Schr\"odinger equation with disorder. This problem is solved exactly on a Bethe lattice. A threshold for delocalization is found to be $\beta_c = 1/\ln 2 \approx 1.4427$. 

For the $\beta$ values smaller than this, the localization persists, despite that the problem is nonlinear. Support for this type of behavior can be found in the results of Ref. \cite{Wang}. In vicinity of the delocalization point the spreading of the wave field is subdiffusive, with second moments that grow with time as a powerlaw $\propto t^\alpha$ for $t\rightarrow+\infty$. We find that $\alpha$ is exactly $1/3$ in the thermodynamic limit. This regime bears signatures enabling to associate it with the onset of ``weak" transport \cite{ChZo} of Alfv\'en eigenmodes in vicinity of marginal stability of magnetic confinement systems. 

The key feature of the critical spreading is that it is completely dominated by multi-scale time correlations that are long-ranged. We associate this regime with Hamiltonian pseudochaos, random non-chaotic dynamics with zero Lyapunov exponents \cite{Report,JMPB,PD2004}. The implications of pseudochaos find their significance in the general picture of transport along separatrices of Hamiltonian systems, the cornerstone of nonlinear dynamics, and are mathematically related with a description in terms of fractional derivative equations. We find it interesting to remark that the above fractional diffusion equation, eq.~(\ref{24}), is ``born" within the mathematical structure of nonlinear Schr\"odinger equation with usual time differentiation. It is in fact the interplay of nonlinearity and randomness that leads to a fractional derivative equation when it comes to non-Markovian transport of the wave field.    

Above criticality, as $\beta$ increases, nonlinearity takes a stronger role over the dynamics. To this end, the multi-scale and pseudochaotic properties turn to lose their importance. For the large $\beta \gg 1$, one may go for a toy-model that singles out the nonlinear properties of the transport. One such model is a Markovian diffusion model with range-dependent diffusion coefficient. In the latter case we find $\alpha = 2/5$ consistently with the results from previous investigations \cite{Sh93,PS}.


\acknowledgments
A.V.M. gratefully acknowledges hospitality at Lewiner Institute for Theoretical Physics at the Technion where this paper was written. This work was supported in part by the Euratom Communities under the contract of Association between EURATOM/ENEA and by the US-Israel Binational Science Foundation (BSF). 



\begin{thebibliography}{0}

\bibitem{And}
P. W. Anderson, Phys. Rev. {\bf 109}, 1492 (1958).

\bibitem{BE}
J. Billy, V. Josse, Z. Zuo, A. Bernard, B. Hambrecht, P. Lugan, D. Cl\'ement, L. Sanchez-Palencia, P. Bouyer and A. Aspect, Nature {\bf 453}, 891 (2008).

\bibitem{Sh93}
D. L. Shepelyansky, Phys. Rev. Lett. {\bf 70}, 1787 (1993).

\bibitem{PS}
A. S. Pikovsky and D. L. Shepelyansky, Phys. Rev. Lett. {\bf 100}, 094101 (2008). 

\bibitem{Flach}
S. Flach, D. O. Krimer, and Ch. Skokos, Phys. Rev. Lett. {\bf 102}, 024101 (2009). 

\bibitem{ChV}
B. V. Chirikov and V. V. Vecheslavov, Zh. \'Eksp. Teor. Fiz. {\bf 112}, 1132 (1997).

\bibitem{Chirikov}
B. V. Chirikov, J. Nucl. Energy Part C: Plasma Phys. {\bf 1}, 253 (1960).

\bibitem{ZaslavskyUFN}
G. M. Zaslavsky and B. V. Chirikov, Phys. Usp. {\bf 14}, 549 (1972); G. M. Zaslavsky, {\it Statistical Irreversibility in Nonlinear Systems} (Nauka, Moscow, 1970).

\bibitem{Sagdeev}
G. M. Zaslavsky and R. Z. Sagdeev, {\it Introduction to the Nonlinear Physics. From Pendulum to Turbulence and Chaos} (Nauka, Moscow, 1988).

\bibitem{PRE01}
A. V. Milovanov, Phys. Rev. E {\bf 63}, 047301 (2001). 

\bibitem{PRE09}
A. V. Milovanov, Phys. Rev. E {\bf 79}, 046403 (2009). 

\bibitem{PRE00}
A. V. Milovanov and G. Zimbardo, Phys. Rev. E {\bf 62}, 250 (2000). 

\bibitem{Report}
G. M. Zaslavsky, Phys. Rep. {\bf 371},  461 (2002).

\bibitem{Klafter}
R. Metzler and J. Klafter, Phys. Rep. {\bf 339}, 1 (2000). 

\bibitem{JMPB}
O. Lyubomudrov, M. Edelman, and G. M. Zaslavsky, Intl. J. Modern Phys. B {\bf 17}, 4149 (2003).  

\bibitem{PD2004}
G. M. Zaslavsky and M. A. Edelman, Physica D {\bf 193}, 128 (2004).

\bibitem{Schroeder}
M. R. Schroeder, {\it Fractals, Chaos, Power Laws: Minutes from an Infinite Paradise} (Freeman, New York, 1991).

\bibitem{Havlin}
S. Havlin and D. ben-Avraham, Adv. Phys. {\bf 51}, 187 (2002); D. ben-Avraham and S. Havlin, {\it Diffusion and Reactions in Fractals and Disordered Systems} (Cambridge University Press, Cambridge, 2000).

\bibitem{Gefen}
Y. Gefen, A. Aharony, and S. Alexander, Phys. Rev. Lett. {\bf 50}, 77 (1983).

\bibitem{Naka}
T. Nakayama, K. Yakubo, and R. L. Orbach, Rev. Mod. Phys. {\bf 66}, 381 (1994).

\bibitem{NJP}
A. V. Milovanov, New J. Phys. {\bf 13}, 043034 (2011). 

\bibitem{Podlubny}
I. Podlubny, {\it Fractional Differential Equations} (Academic Press, San Diego, 1999). 

\bibitem{Nature}
M. F. Shlesinger, G. M. Zaslavsky and J. Klafter, {Nature} {\bf 363}, 31 (1993).

\bibitem{Rest}
R. Metzler and J. Klafter, J. Phys. A: Math. Gen. {\bf 37}, R161 (2004).

\bibitem{Iomin}
A. Iomin, Phys. Rev. E {\bf 81}, 017601 (2010).

\bibitem{Wang}
W.-M. Wang and Z. Zhang, J. Stat. Phys. {\bf 134}, 953 (2009); Y. Krivolapov, S. Fishman, and A. Soffer, New J. Phys. {\bf 12}, 063035 (2010).

\bibitem{ChZo}
F. Zonca, S. Briguglio, L. Chen, G. Fogaccia, and G. Vlad, Nucl. Fusion {\bf 45}, 477 (2005); L. Chen and F. Zonca, ibid. {\bf 47}, S727 (2007).



\end{thebibliography}
\end{document}